# Big Bang Nucleosynthesis

Keith A. Olive[a*]

[a]School of Physics and Astronomy, University of Minnesota,
Minneapolis MN 55455, USA

The concordance of standard big bang nucleosynthesis theory and the related observations of the light element isotopes (including some new higher $^4$He abundances) will be reviewed. Implications of BBN on chemical evolution, dark matter and constraints on particle properties will be discussed.

## 1. Introduction

The question of the concordance of standard big bang nucleosynthesis (BBN) rests on our ability to compare the predictions of the standard model with the available observational data. At this time, as I will argue below, there is indeed concordance between the predictions of BBN and the observational determination of the primordial abundances of $^4$He, $^7$Li, and some measurements of D. Our lack of understanding of the chemical and stellar history of $^3$He prevents us from using this isotope as a critical check of BBN. Recent measurements of the $^4$He mass fraction which indicate a slightly higher abundance, diminish the role of $^4$He as a critical test as the higher abundances are very insensitive to the value of the baryon-to-photon ratio, $\eta$, the key undetermined parameter of BBN. In this case, we must wait until a definitively reliable abundance of D/H (if possible) is available from quasar absorption systems to pin down the value of $\eta$ and definitively check the concordance of BBN.

At this time, it is nevertheless still possible to draw some important inferences from BBN concerning the chemical evolution of the light elements, the amount of non-baryonic dark matter, as well as constraints on particle properties. These issues will be touched upon briefly here. For a more complete discussion see [1].

BBN took place in the early universe when the temperature scale was $\lesssim 1$ MeV. The key events leading to the synthesis of the light elements followed from the period when the weak interaction rates were in equilibrium, thus fixing the ratio of number densities of neutrons to protons. At $T \gg 1$ MeV, $(n/p) \simeq 1$. As the temperature fell and approached the point where the weak interaction rates were no longer fast enough to maintain equilibrium, the neutron to proton ratio was given approximately by the Boltzmann factor, $(n/p) \simeq e^{-\Delta m/T}$, where $\Delta m$ is the neutron-proton mass difference. The final abundance of $^4$He is very sensitive to the $(n/p)$ ratio.

The nucleosynthesis chain begins with the formation of deuterium through the process, $p+n \to D+\gamma$. However, because the large number of photons relative to nucleons, $\eta^{-1} = n_\gamma/n_B \sim 10^{10}$, deuterium production is delayed past the point where the temperature has fallen below the deuterium binding energy, $E_B = 2.2$ MeV (the average photon energy in a blackbody is $\bar{E}_\gamma \simeq 2.7T$). When the quantity $\eta^{-1}\exp(-E_B/T) \sim 1$, the rate for deuterium destruction (D $+\gamma \to p+n$) finally falls below the deuterium production rate and the nuclear chain begins at a temperature

[*]Summary of the talk given at the Vth International Workshop on Theory and Phenomenology in Astroparticle and Underground Physics (TAUP97), Gran Sasso Lab. Italy, September 7-11 1997. This work was supported in part by DOE grant DE-FG02-94ER-40823.

$T \sim 0.1 MeV$.

The dominant product of big bang nucleosynthesis is $^4$He, resulting in an abundance of close to 25% by mass. In fact, it is of key importance that there are no observations of any any system for which the $^4$He mass fraction lies significantly below 25% thus indicating a primordial origin for this element. In the standard model, the $^4$He mass fraction depends primarily on the baryon to photon ratio, $\eta$ as it is this quantity which determines the onset of nucleosynthesis via deuterium production. But because the $(n/p)$ ratio is only weakly dependent on $\eta$, the $^4$He mass fraction is relatively flat as a function of $\eta$. When we go beyond the standard model, the $^4$He abundance is very sensitive to changes in the expansion rate which can be related to the effective number of neutrino flavors as will be discussed below. Lesser amounts of the other light elements are produced: D and $^3$He at the level of about $10^{-5}$ by number, and $^7$Li at the level of $10^{-10}$ by number.

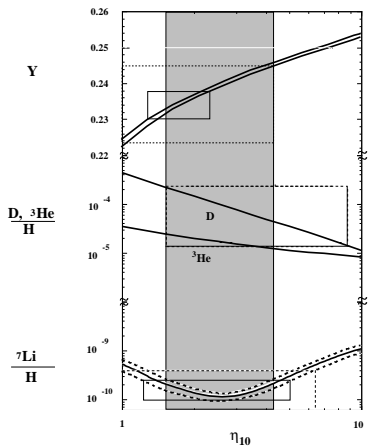

Figure 1. The light element abundances from big bang nucleosynthesis as a function of $\eta_{10} = 10^{10}\eta$.

The resulting abundances of the light elements are shown in Figure 1. The curves for the $^4$He mass fraction, $Y$, bracket the computed range based on the uncertainty of the neutron mean-life which has been taken as [2] $\tau_n = 887 \pm 2$ s. Uncertainties in the produced $^7$Li abundances have been adopted from the results in Hata et al. [3]. Uncertainties in D and $^3$He production are negligible on the scale of this figure. The boxes correspond to the observed abundances and will be discussed below.

## 2. Abundances

### 2.1. $^4$He

There are over 70 different low metallicity extragalactic HII (ionized hydrogen) regions with abundance information on the $^4$He mass fraction, $Y$, O/H, and N/H [4,5,6]. The observation of the heavy elements is important as the helium mass fraction observed in these HII regions has been augmented by some stellar processing, the degree to which is given by the oxygen and nitrogen abundances. In an extensive study based on the data, it was found [7] that the data is well represented by a linear correlation for $Y$ vs. O/H and $Y$ vs. N/H. It is then expected that the primordial abundance of $^4$He can be determined from the intercept of that relation. The $^4$He mass fraction [8] based on 62 distinct HII regions is

$$Y_p = 0.234 \pm 0.002 \pm 0.005 \qquad (1)$$

The first uncertainty is purely statistical and the second uncertainty is an estimate of the systematic uncertainty in the primordial abundance determination. The solid box for $^4$He in Figure 1 represents the range (at $2\sigma_{\rm stat}$) from (1). The dashed box extends this by including $\sigma_{\rm sys}$. A somewhat lower primordial abundance of $Y_p = 0.230 \pm .003$ is found by restricting to the 32 most metal poor regions [8]. A similar result is found using a Bayesian analysis which does not depend on a linear regression of the data [9].

There may be some evidence however, that the $^4$He mass fraction is somewhat higher than in Eq. (1). First, there is a potential problem concerning underlying stellar absorption [10] with one data point (I Zw 18 NW) that is at low O/H and has small errors due its having been observed more than once. When this point is removed, $Y_p$ becomes $0.237 \pm 0.003$ for 61 data points. There is also a new (preliminary) compilation of data from 45 HII regions by Izotov and Thuan [11] which includes new observations as well as the ones published in [6]. Based on their "self-consistent" analysis of 5 helium lines, they find

$Y_p = 0.244 \pm 0.002$, a significantly higher value for $Y_p$. However, some of the lines are very sensitive to the electron density in the HII region. When densities determined from sulphur observations are used (as in [4,5]), they find a lower (but still high) value of $Y_p = 0.239 \pm .002$. This is to be compared with $Y_p = 0.0234 \pm 0.003$ for the data in [7] from [4,5] with I Zw 18 NW removed. When all of the available data is used, I find $Y_p = 0.238 \pm 0.002$ though this should be taken as a tentative value. I will comment below on the effect of higher $Y_p$ on the BBN consistency analysis.

## 2.2. $^7$Li

The $^7$Li abundance is also reasonably well known. In old, hot, population-II stars, $^7$Li is found to have a very nearly uniform abundance. For stars with a surface temperature $T > 5500$ K and a metallicity less than about 1/20th solar (so that effects such as stellar convection may not be important), the abundances show little or no dispersion beyond that which is consistent with the errors of individual measurements. There is $^7$Li data from nearly 100 halo stars, from a variety of sources. I will use the value given in [12] as the best estimate for the mean $^7$Li abundance and its statistical uncertainty in halo stars

$$\text{Li/H} = (1.6 \pm 0.1^{+0.4+1.6}_{-0.3-0.5}) \times 10^{-10} \quad (2)$$

The first error is statistical, and the second is a systematic uncertainty that covers the range of abundances derived by various methods and is probably an overestimate given the analysis in [12]. The solid box for $^7$Li in Figure 1 represents the $2\sigma$ range including this uncertainty. The third set of errors in Eq. (2) accounts for the possibility that as much as half of the primordial $^7$Li has been destroyed in stars, and that as much as 30% of the observed $^7$Li may have been produced in cosmic ray collisions rather than in the big bang. The dashed box in Figure 1, accounts for this additional uncertainty. Observations of $^6$Li, Be, and B help constrain the degree to which these effects play a role [13]. For $^7$Li, the uncertainties are clearly dominated by systematic effects.

## 2.3. D and $^3$He

We have three basic types of abundance information on these two light elements: 1) ISM data, 2) solar system information, and perhaps 3) a primordial abundance of D/H from quasar absorption systems. The best measurement for ISM D/H is [14]

$$(\text{D/H})_{\text{ISM}} = 1.60 \pm 0.09^{+0.05}_{-0.10} \times 10^{-5} \quad (3)$$

The lower bound using this deuterium value establishes an upper bound on $\eta$ which is robust and is shown by the lower right of the solid box in Figure 1. The solar abundance of D/H is inferred from measurements of $^3$He in meteorites and the solar wind. These measurements indicate that [15, 16] $\left(^3\text{He/H}\right)_\odot = (1.5 \pm 0.2 \pm 0.3) \times 10^{-5}$ and $(\text{D/H})_\odot \approx (2.6 \pm 0.6 \pm 1.4) \times 10^{-5}$.

There have been several recent reported measurements of D/H is high redshift quasar absorption systems. Such measurements are in principle capable of determining the primordial value for D/H and hence $\eta$, because of the strong and monotonic dependence of D/H on $\eta$. Though there are several measurements of D/H in such systems, a unique value has yet to be established. The first of these measurements [17] as well as a more recent (relatively low $z$) observation [18] indicate a rather high D/H ratio, D/H $\approx 2 \times 10^{-4}$. Other high D/H ratios were reported in [19]. However, there are reported low values of D/H in other such systems [20] with values D/H $\simeq 2.5 \times 10^{-5}$, significantly lower than the ones quoted above. (The revised value for these is about $3.5 \times 10^{-5}$.) The range of quasar absorber D/H is shown by the dashed box in Figure 1. It is probably premature to use either of these values as the primordial D/H abundance in an analysis of big bang nucleosynthesis, but it is certainly encouraging that future observations may soon yield a firm value for D/H. It is also very unlikely that both are primordial and indicate an inhomogeneity [21] due to microwave background constraints. It is however important to note that there does seem to be a trend that over the history of the Galaxy, the D/H ratio is decreasing, something we expect from galactic chemical evolution. The amount of deuterium astration that has occurred is model dependent.

Finally, there are in addition to the solar $^3$He values, measurements of $^3$He/H in the ISM with [22] $\left(^3\text{He/H}\right)_{\text{ISM}} = 2.1^{+.9}_{-.8} \times 10^{-5}$ and in galactic HII regions [23] which show a wide dispersion which may be indicative of pollution or a bias [24] $\left(^3\text{He/H}\right)_{\text{HII}} \simeq 1 - 5 \times 10^{-5}$. There are also observations of $^3$He in planetary nebulae [25] which show a very high $^3$He abundance of $^3$He/H $\sim 10^{-3}$, corresponding to the production of $^3$He in low mass stars as is expected from standard stellar models [26].

## 3. Likelihood Analysis

Monte Carlo techniques have proven to be a useful form of analysis for big bang nucleosynthesis [27,28,3] and an analysis of this sort was performed [29] concentrating on $^4$He and $^7$Li. It should be noted that in principle, two elements should be sufficient for not only constraining the one parameter ($\eta$) theory of BBN, but also for testing for consistency. Here, I will not describe the procedure for calculating the various likelihood functions but simply show the results. For more information, the reader is referred to [29].

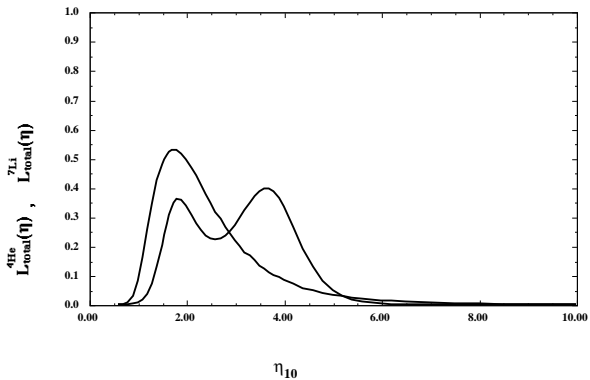

Figure 2. Likelihood distribution for each of $^4$He and $^7$Li, shown as a function of $\eta$.

What are plotted in Figure 2 are the individual likelihood functions for $^4$He and $^7$Li as a function of $\eta_{10}$. The assumed observational element abundances were taken from Eqs. (1) and (2). The one-peak structure of the $^4$He curve corresponds to its monotonic increase with $\eta$, while the two-peaks for $^7$Li arise from its passing through a minimum ( i.e., a given observed value of $^7$Li, there are two likely values of $\eta$) as can be surmised from Figure 1. As one can see there is very good agreement between $^4$He and $^7$Li in the vicinity of $\eta_{10} \simeq 1.8$.

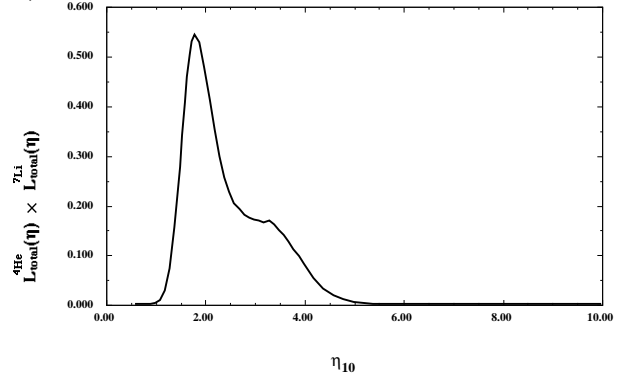

Figure 3. Combined likelihood for simultaneously fitting $^4$He and $^7$Li, as a function of $\eta$.

The combined likelihood, for fitting both elements simultaneously, is given by the product of the two functions in Figure 2 and is shown in Figure 3. The likelihood prediction (95% CL) for $\eta$ is $\eta_{10} = 1.8^{+2.4}_{-.4}$ corresponding to

$$\Omega_B h^2 = .006^{+.009}_{-.001} \qquad (4)$$

where $\Omega_B$ is the fraction of critical density in baryons, and $h$ is the Hubble parameter scaled to 100 km/s/Mpc.

It is interesting to note that the central (and strongly) peaked value of $\eta_{10}$ determined from the combined $^4$He and $^7$Li likelihoods corresponds to a value of D/H $\simeq 1.8 \times 10^{-4}$, very close [30] to the high value of D/H in quasar absorbers [17,18,19]. This is one of the main motivations for studying galactic chemical evolution models with high initial D/H and strong D destruction histories.

The effects of the higher values of $Y_p$ discussed in section 2.1 can be seen in Figure 4. Shown as solid curves are the same two likelihood functions from Figure 2 for $^4$He and $^7$Li. Also shown in solid are the likelihood functions for high D/H (at low $\eta_{10}$) and low D/H (at high $\eta_{10}$). As can be seen, there is striking agreement between $^4$He, $^7$Li, and high D/H. The dotted curve corresponds to $Y_p = 0.238 \pm 0.002$ and as one can see there is still broad agreement with $^7$Li and high D/H. The maximum likelihood value for $\eta$ is $\eta_{10} = 1.9$ with a 95% CL range of $1.6 < \eta_{10} < 4.6$. The dashed curve shows the result for $Y_p = 0.244 \pm 0.002$. In this case, we learn very little from $^4$He as its likelihood function is now spread over nearly all values of $\eta$ in the region of interest. There is overlap between this distribution and $^7$Li (though it now favors the high $\eta$ peak) as well as overlap with *both* high and low D/H. The maximum likelihood value for $\eta$ in this case is $\eta_{10} = 3.7$ with a 95% CL range of $1.8 < \eta_{10} < 5.0$ from both $^4$He and $^7$Li. For completeness, I note that the 95% CL range for high D/H is $1.3 < \eta_{10} < 3.8$ and for low D/H it is $4.8 < \eta_{10} < 6.2$.

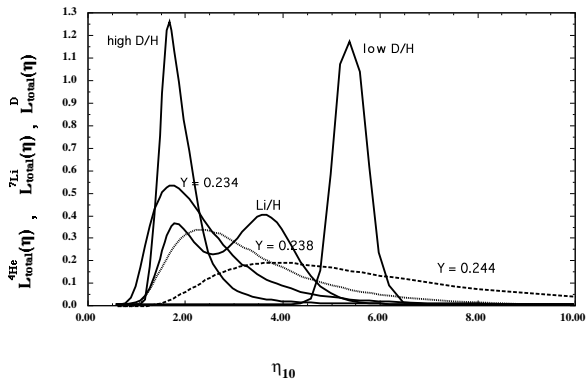

Figure 4. A comparison of likelihood distributions for $^4$He with $Y_p = 0.234, 0.238$, and $0.244$ and $^7$Li, as well as high and low D/H.

## 4. Implications

### 4.1. Chemical Evolution

Because we can not directly measure the primordial abundances of any of the light element isotopes, we are required to make some assumptions concerning the evolution of these isotopes. As has been discussed above, $^4$He is produced in stars along with oxygen and nitrogen. $^7$Li can be destroyed in stars and produced in several (though still uncertain) environments. D is totally destroyed in the star formation process and $^3$He is both produced and destroyed in stars with fairly uncertain yields. It is therefore preferable, if possible to observe the light element isotopes in a low metallicity environment. Such is the case with $^4$He and $^7$Li, and we can be fairly assured that the abundance determinations of these isotopes are close to primordial. If the quasar absorption system measurements of D/H stabilize, then this too may be very close to a primordial measurement. Otherwise, to match the solar and present abundances of D and $^3$He to their primordial values requires a model of galactic chemical evolution.

It is beyond the scope of this discussion to review galactic chemical evolution, but some comments are in order. Deuterium is always a monotonically decreasing function of time in chemical evolution models. The degree to which D is destroyed, is however a model dependent question. The evolution of $^3$He is considerably more complicated. Stellar models predict that substantial amounts of $^3$He are produced in stars between 1 and 3 $M_\odot$ and is consistent with the observation of high $^3$He/H in planetary nebulae [25]. However, the inclusion of the standard $^3$He yield in chemical evolution models leads to an overproduction of $^3$He/H particularly at the solar epoch [24,31]. This problem is compounded in models with an intense period of D destruction. In Scully et al. [32], a dynamically generated supernovae wind model was coupled to models of galactic chemical evolution with the aim of reducing a primordial D/H abundance of $2 \times 10^{-4}$ to the present ISM value without overproducing heavy elements and remaining within the other observational constraints typically imposed on such models. The time evolution of D/H and $^3$He/H from this model is shown in Figure 5. The dashed curve assumes standard yields for $^3$He.

The overproduction of $^3$He relative to the solar meteoritic value seems to be a generic feature of chemical evolution models when $^3$He production in low mass stars is included. This conclusion is independent of the chemical evolution model and is directly related to the assumed stellar yields of $^3$He. It has recently been suggested that at least some low mass stars may indeed be net destroyers of $^3$He if one includes the effects of rotational mixing in low mass stars on the red giant branch [33,34]. It was shown [35] that these reduced $^3$He yields in low mass stars can account for the relatively low solar and present day $^3$He/H abundances observed as shown by the solid $^3$He curve in Figure 5.

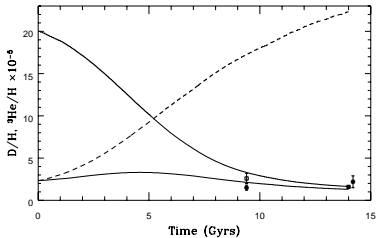

Figure 5. The evolution of D/H and $^3$He/H as a function of time.

## 4.2. Dark Matter

The implications of the resulting predictions from big bang nucleosynthesis on dark matter are clear. First, if $\Omega = 1$ (as predicted by inflation), and $\Omega_B \lesssim 0.1$ which is certainly a robust conclusion based on D/H, then non-baryonic dark matter is a necessity. Second, on the scale of small groups of galaxies, $\Omega \gtrsim 0.05$, and is expected to sample the dark matter in galactic halos. This value is probably larger than the best estimate for $\Omega_B$ from equation (4). $\Omega_B h^2 = 0.006$ corresponds to $\Omega_B \simeq 0.025$ for $h = 1/2$. In this event, some non-baryonic dark matter in galactic halos is required. This conclusion is unchanged by the inclusion of the high D/H measurements in QSO absorbers. In contrast, the low D/H measurements would imply that $\Omega_B h^2 = 0.023$ allowing for the possibility that $\Omega_B \simeq 0.1$. In this case, no non-baryonic dark matter is required in galactic halos.

## 4.3. Particle Properties

Limits on particle physics beyond the standard model from BBN are mostly sensitive to the bounds imposed on the $^4$He abundance. As is well known, the $^4$He abundance is predominantly determined by the neutron-to-proton ratio just prior to nucleosynthesis and is easily estimated assuming that all neutrons are incorporated into $^4$He. As discussed earlier, the neutron-to-proton ratio is fixed by its equilibrium value at the freeze-out of the weak interaction rates at a temperature $T_f \sim 1$ MeV modulo the occasional free neutron decay. Freeze-out is determined by the competition between the weak interaction rates and the expansion rate of the Universe

$$G_F^2 T_f^5 \sim \Gamma_{\text{wk}}(T_f) = H(T_f) \sim \sqrt{G_N N} T_f^2 \quad (5)$$

where $N$ counts the total (equivalent) number of relativistic particle species. The limit on $N_\nu$ comes about via the change in the expansion rate given by the Hubble parameter,

$$H^2 = \frac{8\pi G}{3}\rho = \frac{8\pi^3 G}{90}[N_{\text{SM}} + \frac{7}{8}\Delta N_\nu]T^4 \quad (6)$$

when compared to the weak interaction rates. Here $N_{\text{SM}}$ refers to the standard model value for N. At $T \sim 1$ MeV, $N_{\text{SM}} = 43/4$. The presence of additional neutrino flavors (or any other relativistic species) at the time of nucleosynthesis increases the overall energy density of the Universe and hence the expansion rate leading to a larger value of $T_f$, $(n/p)$, and ultimately $Y_p$. Because of the form of Eq. (5) it is clear that similar to placing limits [36] on $N$, any changes in the weak or gravitational coupling constants can be also constrained (for a recent discussion see [37]).

Just as $^4$He and $^7$Li were sufficient to determine a value for $\eta$, a limit on $N_\nu$ can be obtained as well [29,38]. Since the light element abundances can be computed as functions of both $\eta$ and $N_\nu$, the likelihood method can be extended to include both $\eta$ and $N_\nu$ as arguments. A three-dimensional view of the combined likelihood functions [38] is shown in Figure 6. In this case the high and low $\eta$ maxima of Figure 2, show up as peaks in the $L - \eta - N_\nu$ space. The peaks of the distribution as well as the allowed ranges of $\eta$ and $N_\nu$ are more easily discerned in the contour plot of Figure 7 which shows the 50%, 68% and 95% confidence level contours in the likelihood function. The crosses show the location of the peaks of the likelihood functions. $L_{47}$ peaks at $N_\nu = 3.0$, $\eta_{10} = 1.8$ (in agreement with our previous results [29]) and at $N_\nu = 2.3$, $\eta_{10} = 3.6$. The 95% confidence level allows the following ranges in $\eta$ and $N_\nu$

$$1.6 \leq N_\nu \leq 4.0 \qquad 1.3 \leq \eta_{10} \leq 5.0 \qquad (7)$$

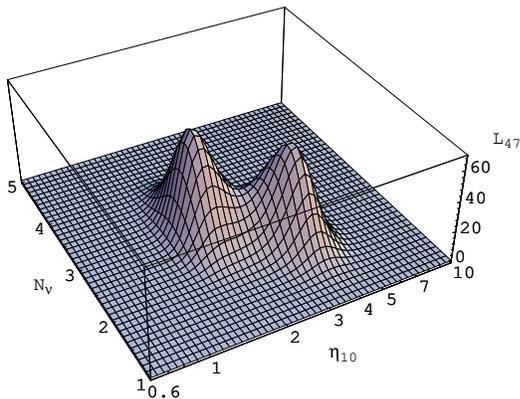

Figure 6. The combined two-dimensional likelihood function, $L_{47}$ for simultaneously fitting $^4$He and $^7$Li as functions of both $\eta$ and $N_\nu$.

Since it is the $^4$He abundance which is most sensitive to $N_\nu$, one might expect that the limit on $N_\nu$ depends on our choice of $Y_p$. It turns out however, that the upper limit to $N_\nu$ is relatively firm. For $Y_p = 0.238 \pm 0.002$, the peak value of $\eta_{10} = 1.9$ corresponds to a peak value $N_\nu = 3.2$ and a 95% CL upper limit, $N_\nu < 4.1$. Similarly, for $Y_p = 0.244 \pm 0.002$, the peak value of $\eta_{10} = 3.7$ corresponds to a peak value $N_\nu = 3.0$ and a 95% CL upper limit, $N_\nu < 4.0$. Although the value of $Y_p$ is moving up allowing for more neutrinos, the peak value of $\eta$ is also moving up and compensates by allowing fewer neutrino flavors. The net result is a robust limit.

One should recall that the limit derived above is not meant for neutrinos in the strictest sense. That is, the limit is only useful when applied to additional particle degrees of freedom which necessarily do not couple to the $Z^o$. For very weakly interacting particles, one must take into account the reduced abundance of these particles at the time of nucleosynthesis[39]. For example, when the limit is applied to scalar fields we find that a single additional scalar degree of freedom (which counts as $\frac{4}{7}$) such as a majoron is allowed. On the other hand, in models with right-handed interactions, and three right-handed neutrinos, the constraint is severe. The right-handed states must have decoupled early enough to ensure that their present temperature have the ratio $(T_{\nu_R}/T_{\nu_L})^4 < 1/3$. If right-handed interactions are mediated by additional gauge interactions, associated with some scale $M_{Z'}$, and if we assume that the right handed interactions scale as $M_{Z'}^4$ with a standard model-like coupling, then it is possible to set a limit on the mass scale $M_{Z'} \gtrsim 1.6$ TeV! In general this constraint is very sensitive to the BBN limit on $N_\nu$. Of course BBN can be used to constrain much more than the number of neutrino flavors and the reader is referred to [40] for a more complete review.

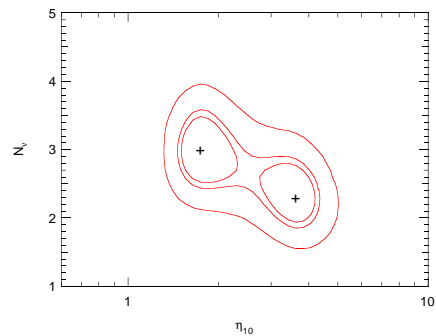

Figure 7. Likelihood contours in the combined likelihood function for $^4$He and $^7$Li. The contours represent 50% (innermost), 68% and 95% (outermost) confidence levels. The crosses mark the points of maximum likelihood.